\documentclass[showpacs,showkeys,byrevtex,twocolumn,pra]{revtex4}%
\usepackage{amsfonts}
\usepackage{amsmath}
\usepackage{amssymb}
\usepackage{graphicx}
\usepackage{verbatim}
\usepackage{amsthm}%
\providecommand{\U}[1]{\protect\rule{.1in}{.1in}}
\renewenvironment{proof}[1][Proof]{\noindent\textbf{#1.} }{\ \rule{0.5em}{0.5em}}

\newtheorem*{proposition}{Proposition}

\def\U{{\hat U}}

\def\bea{\begin {eqnarray*}}
\def\eea{\end {eqnarray*}}

\def\bar{\overline}
\def\*{\star}
\def\({\left(}

\def\){\right)}

\def\2pi{\hbox{$2\pi i$}}

\def\dsl{\raise.15ex\hbox{/}\kern-.57em\partial}
\def\Dsl{\,\raise.15ex\hbox{/}\mkern-.13.5mu D}

\def\<{\langle}
\def\>{\rangle}

\def\CH{{\cal H}}

\def\1{{\mathbf{1} }}

\def\2pi{\hbox{$2\pi i$}}

\def\dsl{\raise.15ex\hbox{/}\kern-.57em\partial}
\def\Dsl{\,\raise.15ex\hbox{/}\mkern-.13.5mu D}
\def\beq{\begin {equation}}
\def\eeq{\end {equation}}

\begin{document}
\preprint{ }
\title[ ]{Encoding One Logical Qubit Into Six Physical Qubits}
\author{Bilal Shaw$^{1,4,5}$}
\email{bilalsha@usc.edu}
\author{Mark M. Wilde$^{1,5}$}
\author{Ognyan Oreshkov$^{2,5}$}
\author{Isaac Kremsky$^{2,5}$}
\author{Daniel A. Lidar$^{1,2,3,5}$}
\affiliation{$^{1}$Department of Electrical
Engineering, $^{2}$Department of Physics and Astronomy, $^{3}$Department of
Chemistry, $^{4}$Department of Computer Science,
$^{5}$Center for Quantum Information Science and Technology, University of Southern California, Los Angeles, California 90089, USA}
\begin{abstract}
We discuss two methods to encode one qubit into six physical
qubits. Each of our two examples corrects an arbitrary
single-qubit error. Our first example is a degenerate six-qubit
quantum error-correcting code. We explicitly provide the
stabilizer generators, encoding circuit, codewords, logical Pauli
operators, and logical CNOT operator for this code. We also show
how to convert this code into a non-trivial subsystem code
that saturates the subsystem Singleton bound.  We then prove that a
six-qubit code without entanglement assistance cannot
simultaneously possess a Calderbank-Shor-Steane (CSS) stabilizer
and correct an arbitrary single-qubit error.  A corollary of this
result is that the Steane seven-qubit code is the smallest
single-error correcting CSS code.  Our second example is the construction of
a non-degenerate six-qubit CSS entanglement-assisted code. This code uses one bit of
entanglement (an ebit) shared between the sender and the receiver and
corrects an arbitrary single-qubit error. The code we obtain
is globally equivalent to the Steane seven-qubit code and
thus corrects an arbitrary error on the receiver's half of the ebit as well.
We prove that this code is the smallest code with a CSS structure that
uses only one ebit and corrects an arbitrary single-qubit error on
the sender's side. We discuss the advantages and disadvantages for
each of the two codes.
\end{abstract}
\volumeyear{2008}
\volumenumber{ }
\issuenumber{ }
\eid{ }
\date{\today}
\received{\today}

\revised{}

\accepted{}

\published{}

\startpage{1}
\endpage{ }
\maketitle

\section{Introduction}

It has been more than a decade since Peter Shor's seminal paper on quantum
error correction \cite{PRA.52.R2493.1995}. He showed how to protect one qubit
against decoherence by encoding it into a subspace of a Hilbert space larger
than its own. For the first time, it was possible to think about quantum
computation from a practical standpoint.

Calderbank and Shor then provided asymptotic rates for the
existence of quantum error-correcting codes and gave upper bounds
for such rates \cite{PRA.54.1098.1996}. They defined a quantum
error-correcting code as an isometric map that encodes $k$ qubits
into a subspace of the Hilbert space of $n$ qubits. As long as
only $t$ or fewer qubits in the encoded state undergo errors, we
can decode the state correctly. The notation for describing such
codes is $[[n,k,d]]$, where $d$ represents the distance of
the code, and the code encodes $k$ logical qubits into $n$
physical qubits.

These earlier codes are examples of additive or stabilizer codes. Additive codes
encode quantum information into the +1
eigenstates of $n$-fold tensor products of Pauli operators
\cite{PRA.54.1862.1996,thesis97gottesman}. Gottesman developed an elegant
theory, the stabilizer formalism, that describes error correction, detection,
and recovery in terms of algebraic group theory.

Steane constructed a seven-qubit code that encodes one qubit, corrects an arbitrary
single-qubit error, and is an example of a Calderbank-Shor-Steane (CSS)\ code
\cite{Steane:96a}.  The five-qubit quantum error-correcting code is a \textquotedblleft perfect
code\textquotedblright\ in the sense that it encodes one qubit with the
smallest number of physical qubits while still correcting an arbitrary single-qubit
error \cite{PRL.77.198.1996,PRA.54.3824.1996}.

Even though every stabilizer code is useful for fault-tolerant
computation \cite{PRA.54.1862.1996,thesis97gottesman}, CSS codes
allow for simpler fault-tolerant procedures. For example, an
encoded CNOT gate admits a transversal implementation without the
use of ancillas if and only if the code is of the CSS type
\cite{thesis97gottesman}. The five-qubit code is not a CSS code
and does not possess the simple fault-tolerant properties of CSS
codes \cite{book2000mikeandike}. The Steane code is a CSS code and
is well-suited for fault-tolerant computation because it has
bitwise implementations of the Hadamard and the phase gates as
well (the logical $X$ and $Z$ operators have bitwise
implementations for any stabilizer code \cite{PRA.54.1862.1996}).
However, an experimental realization of the seven-qubit code may
be more difficult to achieve than one for the five-qubit code
because it uses two additional physical qubits for encoding.

Calderbank {\it et al.}~discovered two distinct six-qubit quantum
codes \cite{ieee1998calderbank} which encode one qubit and correct
an arbitrary single-qubit error. They discovered the first of
these codes by trivially extending the five-qubit code and the other one
through an exhaustive search of the encoding space. Neither of
these codes is a CSS code.

The five-qubit code and the Steane code have been studied
extensively \cite{book2000mikeandike}, but the possibility for encoding one
qubit into six has not received much attention except for the
brief mention in Ref.~\cite{ieee1998calderbank}. In the current paper, we
bridge the gap between the five-qubit code and the Steane code by
discussing two examples of a six-qubit code. The first code we present is a standard
stabilizer code and the second is an entanglement-assisted code. We have not
explicitly checked whether our first example
is equivalent to the non-trivial code of Calderbank~{\it et al}., but we provide
a logical argument in a subsequent paragraph to show that they are equivalent. We also present
several proofs concerning the existence of single-error-correcting
CSS codes of a certain size. One of our proofs gives insight into
why Calderbank {\it et al.}~were unable to find a six-qubit CSS
code. The other proofs use a technique similar to the first proof
to show the non-existence of a CSS entanglement-assisted code that
uses fewer than six local physical qubits where one of the local qubits is half of one ebit, and corrects
an arbitrary single-qubit error.

We structure our work according to our four main results. We first
present a degenerate six-qubit quantum code and show how to
convert this code to a subsystem code. Our second result is a
proof for the non-existence of a single-error-correcting CSS
six-qubit code. Our third result is the construction of a
six-qubit CSS entanglement-assisted quantum code. This code is
globally equivalent to the Steane code. We finally
show that the latter is the smallest example of an
entanglement-assisted CSS code that corrects an arbitrary
single-qubit error.

In Section~\ref{sec:isaac}, we present a degenerate
six-qubit quantum error-correcting code that corrects an
arbitrary single-qubit error.  We present the logical Pauli operators
, CNOT and encoding circuit for this code.  We also prove
that a variation of this code gives us a non-trivial example of
a subsystem code that saturates the
subsystem Singleton bound~\cite{klap0703213}.

In Section~\ref{sec:ogy}, we present a proof that a single-error-correcting CSS six-qubit
code does not exist.
Our proof enumerates all possible CSS forms for the five
stabilizer generators of the six-qubit code and shows that none of
these forms corrects the set of all single-qubit errors.

Section~\ref{sec:bilal-mark} describes the construction of a
six-qubit non-degenerate entanglement-assisted CSS code and
presents its stabilizer generators, encoding circuit, and logical Pauli operators. This code
encodes one logical qubit into six local physical qubits. One of
the physical qubits used for encoding is half of an ebit that the
sender shares with the receiver. The six-qubit
entanglement-assisted code is globally equivalent to
the seven-qubit Steane code~\cite{Steane:96a} and thus corrects an
arbitrary single-qubit error on all of the qubits (including the
receiver's half of the ebit). This ability to correct errors on
the receiver's qubits in addition to the sender's qubits is not
the usual case with codes in the entanglement-assisted paradigm, a
model that assumes the receiver's halves of the ebits are noise
free because they are already on the receiving end of the channel.
We show that our example is a trivial case of a more general
rule---every $[[n,1,3]]$ code is equivalent to a $[[n-1,1,3;1]]$
entanglement-assisted code by using any qubit as Bob's half of the
ebit.

Finally, in section~\ref{sec:ogycss}, we present a proof that the
Steane code is an example of the smallest entanglement-assisted
code that corrects an arbitrary single-qubit error on the sender's
qubits, uses only one ebit, and possesses the CSS form.

The appendix gives a procedure to obtain the encoding circuit for
the six-qubit CSS entanglement-assisted code. It also lists a
table detailing the error-correcting properties for the
degenerate six-qubit code.

\section{Degenerate Six-Qubit Quantum Code}

\label{sec:isaac}This section details an example of a six-qubit
code that corrects an arbitrary single-qubit error.  We explicitly present the
stabilizer generators, encoding circuit, logical codewords,
logical Pauli operators and CNOT\ operator for this
code. We also show how to convert this code into a
subsystem code where one of the qubits is a gauge qubit.  We
finish this section by discussing the advantages and disadvantages
of this code.

Calderbank {\it et al.} mention the existence of two non-equivalent six-qubit codes~\cite{ieee1998calderbank}.
Their first example is a trivial extension of the five-qubit code.
They append an ancilla qubit to the five-qubit code to obtain this code.
Their second example is a non-trivial six-qubit code. They argue that there are no other codes ``up to equivalence.''
Our example is not reducible to the trivial six-qubit code because every one of its qubits is entangled with the others.
It therefore is equivalent to the second non-trivial six-qubit code in Ref.~\cite{ieee1998calderbank}
according to the arguments of Calderbank  {\it et al.}

\begin{table}[tbp] \centering
\begin{tabular}
[c]{c|cccccc}\hline\hline
$h_{1}$ & $Y$ & $I$ & $Z$ & $X$ & $X$ & $Y$\\
$h_{2}$ & $Z$ & $X$ & $I$ & $I$ & $X$ & $Z$\\
$h_{3}$ & $I$ & $Z$ & $X$ & $X$ & $X$ & $X$\\
$h_{4}$ & $I$ & $I$ & $I$ & $Z$ & $I$ & $Z$\\
$h_{5}$ & $Z$ & $Z$ & $Z$ & $I$ & $Z$ & $I$\\ \hline
$\overline{X}$ & $Z$ & $I$ & $X$ & $I$ & $X$ & $I$\\
$\overline{Z}$ & $I$ & $Z$ & $I$ & $I$ & $Z$ & $Z$
\\\hline\hline
\end{tabular}
\caption{Stabilizer generators $h_1$, \ldots, $h_5$, and logical operators $\bar{X}$ and $\bar{Z}$ for the six-qubit code.  The convention in the above generators is that $Y=ZX$.}\label{tbl:isaac-613}
\end{table}%

\begin{figure}
[ptb]
\begin{center}
\includegraphics{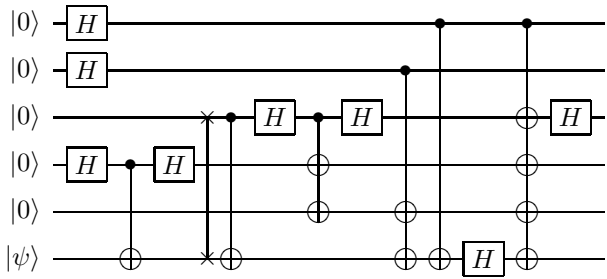}
\end{center}
\caption
{Encoding circuit for the first six-qubit code.  The \textit{H} gate is a Hadamard gate.  For example, we apply
a Hadamard on qubit four followed by a CNOT with qubit four as the control qubit and qubit six as the target qubit. }
\label{fig:six-qubit-1}
\end{figure}

Five generators specify the degenerate six-qubit code.
Table~\ref{tbl:isaac-613} lists the generators $h_{1}$, \ldots, $h_{5}$ in the stabilizer $\mathcal{S}$, and the logical operators $\overline{X}$ and
$\overline{Z}$ for the six-qubit code. Figure~\ref{fig:six-qubit-1}\ illustrates an
encoding circuit for the six-qubit code. The encoding circuit is not fault tolerant, but one 
can consult Refs.~\cite{PRA.54.1862.1996,thesis97gottesman} to determine fault-tolerant procedures for arbitrary stabilizer codes.

The quantum error-correcting conditions guarantee that the six-qubit
code corrects an arbitrary single-qubit error
\cite{book2000mikeandike}. Specifically, the error-correcting
conditions are as follows: a stabilizer $\mathcal{S}$ with
generators $s_i$ where $i=1,\ldots,n-k$ (in our case $n=6$ and
$k=1$), corrects an error set $\mathcal{E}$ if every error pair
$E^{\dag}_{a}E_{b}\in \mathcal{E}$ either anticommutes with at
least one stabilizer generator
\begin{equation}
\exists\ s_{i}\in\mathcal{S}:\left\{  s_{i},E_{a}^{\dag}E_{b}\right\}  =0,
\end{equation}
or is in the stabilizer,
\begin{equation}
E_{a}^{\dag}E_{b}\in\mathcal{S}.
\end{equation}
These conditions imply the ability to correct any linear
combination of errors in the set $\mathcal{E}$
\cite{book2000mikeandike,book2007mermin}.  At least one
generator from the six-qubit stabilizer anticommutes with each of the
single-qubit Pauli errors, $X_{i},Y_{i},Z_{i}$ where
$i=1,\ldots,6$, because the generators have at least one
$Z$ and one $X$ operator in all six positions. Additionally, at
least one generator from the stabilizer anticommutes with each
pair of two distinct Pauli errors (except $Z_{4}Z_{6}$, which is
in the stabilizer $\mathcal{S}$).  Table~\ref{table:sixonethree1} in the 
appendix lists such a
generator for every pair of distinct Pauli errors for the
six-qubit code. These arguments and the table listings prove that
the code can correct an arbitrary single-qubit error.

The logical basis states for the six-qubit code are as follows:
\begin{align*}
|\overline{0}\rangle &  =%
\begin{array}
[c]{c}%
|000000\rangle-|100111\rangle+|001111\rangle-|101000\rangle\ -\\
|010010\rangle+|110101\rangle+|011101\rangle-|111010\rangle
\end{array},\\
|\overline{1}\rangle &  =%
\begin{array}
[c]{c}%
|001010\rangle+|101101\rangle+|000101\rangle+|100010\rangle\ -\\
|011000\rangle-|111111\rangle+|010111\rangle+|110000\rangle
\end{array},
\end{align*}
where we suppress the normalization factors of the above codewords. 

\begin{figure}
[ptb]
\begin{center}
\includegraphics{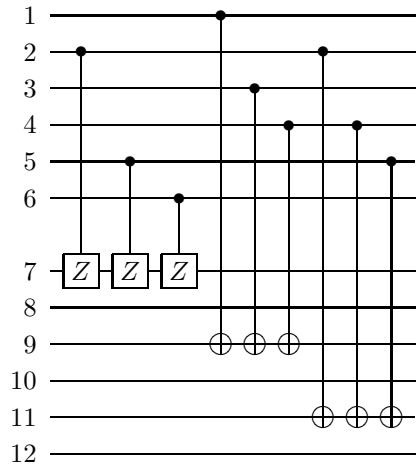}
\end{center}
\caption{Logical CNOT for the six-qubit quantum code.  The first six qubits represent a logical source qubit and the last six represent a logical target qubit.  For example we begin the circuit by applying a CZ (controlled-Z) gate from source qubit two to target qubit seven.}\label{fig-logicalcnot-1}
\end{figure}

A series of CNOT and controlled-$Z$ operations implement the logical CNOT
operation for the six-qubit code.  Let CN$\left(  i,j\right)  $ denote a CNOT
acting on physical qubits $i$ and $j$ with qubit $i$ as
the control and qubit $j$ as the target. Let CZ$\left(  i,j\right)  $
denote controlled-$Z$ operations.\ The logical CNOT for the six-qubit code is as follows:%
\begin{multline*}
\overline{\text{CNOT}}=\text{CZ}\left(  2,7\right)  \ \text{CZ}\left(
5,7\right)  \ \text{CZ}\left(  6,7\right)  \ \text{CN}\left(  1,9\right)  \\
\text{CN}\left(  3,9\right)  \ \text{CN}\left(  4,9\right)  \ \text{CN}%
\left(  2,11\right) \ \text{CN}\left(  4,11\right)  \ \text{CN}\left(  5,11\right)
\end{multline*}
Figure~\ref{fig-logicalcnot-1} depicts the logical CNOT\ acting on two logical qubits
encoded with the six-qubit code. 

Both the six-qubit code and the five-qubit code correct
an arbitrary single-qubit error. But the six-qubit code has the
advantage that it corrects a larger set of errors than the
five-qubit code. This error-correcting capability comes at the expense of a larger
number of qubits---it corrects a larger set of errors because the Hilbert
space for encoding is larger than that for the five-qubit code. In
comparison to the Steane code, the six-qubit code uses a smaller
number of qubits, but the disadvantage is that it does not admit
a simple transversal implementation of the logical CNOT. In
addition, the Steane code admits a bitwise implementation
of all logical single-qubit Clifford gates whereas the six-qubit
code does not.

\subsection{Subsystem Code Construction}

\label{subsystem}We convert the degenerate six-qubit code from the previous
section into a subsystem code. The degeneracy inherent in the code allows us
to perform this conversion.  The code
still corrects an arbitrary single-qubit error after we replace one of the
unencoded ancilla qubits with a gauge qubit. 

We briefly review the history of subsystem codes. The essential
insight of Knill {\it et al.} was that the most general way to
encode quantum information is into a subsystem rather than
a subspace \cite{knillPRL00}. In the case when the information is
encoded in a single subsystem, the Hilbert space decomposes as
$\CH = (\CH_{A} \otimes \CH_{B})\oplus \CH_{C}$ where the
subsystem $\CH_{A}$ stores the protected information. Errors that
act on subsystem $\CH_{B}$, also known as the gauge subsystem, do
not require active correction because $\CH_{B}$ does not store any
valuable information. This passive error-correction ability of a
subsystem code may lead to a smaller number of stabilizer
measurements during the recovery process and may lead to an
improvement of the accuracy threshold for quantum computation \cite{aliferisPRL07}.
Kribs {\it et al.} recognized in Ref.~\cite{kribsPRL05} that
this subsystem structure of a Hilbert space is useful
for active quantum error-correction as well (Knill {\it et al.} did
not explicitly recognize this ability in Ref.~\cite{knillPRL00}.)
See Ref.~\cite{isit2008aly} for a discussion of all aspects of subsystem
code constructions and a detailed theoretical comparison
between subsystem and stabilizer codes.

We now detail how to convert the six-qubit code from the
previous section into a subsystem code.  The sixth unencoded qubit is the
information qubit and the encoding operation transforms it into
subsystem $\CH_A$. We convert the fourth unencoded ancilla qubit to a
gauge qubit. We simply consider it as a noisy qubit so that the
operators $X_4$ and $Z_4$ have no effect on the quantum
information stored in subsystem $\CH_{A}$. The operators $X_4$ and
$Z_4$ generate the unencoded gauge group. The encoding circuit in
Figure~\ref{fig:six-qubit-1} transforms these unencoded operators
into $X_4$ and $Z_4Z_6$ respectively. These operators together
generate the encoded gauge subgroup $H = \left\langle
X_4,Z_4Z_6\right\rangle$.  Errors in this subgroup do not affect the
encoded quantum information.  The code is still able to correct an
arbitrary single-qubit error because each one of the single-qubit
Pauli error pairs anticommutes with at least one of the generators
from the new stabilizer $\widetilde{\mathcal{S}}=\left\langle
h_1,h_2,h_3,h_5\right\rangle$, or belong to $H$
\cite{poulinPRL05}. Table~\ref{table:sixonethree1} shows this
property for all error pairs. The code passively corrects the error pairs $X_4$, $Z_4Z_6$,
$Y_4Z_6$ because they belong to the gauge subgroup.

\begin{table}[tbp] \centering
\begin{tabular}
[c]{c|cccccc}\hline\hline
$h_{1}$ & $Y$ & $I$ & $Z$ & $X$ & $X$ & $Y$\\
$h_{2}$ & $Z$ & $X$ & $I$ & $I$ & $X$ & $Z$\\
$h_{3}$ & $I$ & $Z$ & $X$ & $X$ & $X$ & $X$\\
$h_{5}$ & $Z$ & $Z$ & $Z$ & $I$ & $Z$ & $I$\\ \hline
$H_{X}$ & $I$ & $I$ & $I$ & $X$ & $I$ & $I$\\
$H_{Z}$ & $I$ & $I$ & $I$ & $Z$ & $I$ & $Z$\\ \hline
$\overline{X}$ & $Z$ & $I$ & $X$ & $I$ & $X$ & $I$\\
$\overline{Z}$ & $I$ & $Z$ & $I$ & $I$ & $Z$ & $Z$
\\\hline\hline
\end{tabular}
\caption{Stabilizer generators $h_1, h_2, h_3$ and $h_5$, gauge subgroup generators $H_{X}$ and $H_{Z}$, and logical operators $\bar{X}$ and $\bar{Z}$ for the six-qubit code.  The convention in the above generators is that $Y=ZX$.}\label{tbl:subsystem}%
\end{table}%

The six-qubit single-error-correcting subsystem code
discussed above saturates the Singleton bound for subsystem codes
\cite{klap0703213},
\begin{equation}
n -k - r \geq 2(d - 1),
\end{equation}
where for our code, $n=6$, $k=1$, $r=1$, and $d=3$.
This code is the smallest non-trivial subsystem
code that corrects an arbitrary single-qubit
error and is a code that satisfies the conjecture at the end of Ref.~\cite{arx2008aly}. A trivial way to saturate this bound is to add a noisy qubit to the five-qubit code!  One of
the advantages of using the subsystem construction is that we only
need to perform four stabilizer measurements instead of five
during the recovery process.

\section{Non-existence of a $[[6,1,3]]$ CSS Code}

\label{sec:ogy}Our proposition below proves that it is impossible
for a six-qubit code to possess the CSS\ structure while
correcting an arbitrary single-qubit error.  An immediate
corollary of this proposition is that the seven-qubit code is the smallest
single-error-correcting CSS code.

\begin{proposition}
There is no six-qubit code that encodes one qubit, possesses the CSS\ structure, and corrects an
arbitrary single-qubit error.
\end{proposition}

\begin{proof}
We first suppose that a code with the above properties exists. If
a $[[6,1,3]]$ CSS code exists, its stabilizer $\mathcal{S}$ must
have five generators:
\begin{equation}
\mathcal{S}=\langle g_{1},...,g_{5}\rangle.
\end{equation}
The CSS\ structure implies that each of these generators includes
$X$ operators only or $Z$ operators only (except for the
identity). The set of correctable Pauli errors $\{E_{j}\}$ in the
Pauli group acting on six qubits satisfies
$\{E_{i}E_{j},\mathcal{S}\}=0$ unless $E_{i}E_{j}\in \mathcal{S}$,
for all $i,j$. We show below that no set of five CSS\ stabilizer
generators acting on six qubits can correct an arbitrary
single-qubit error and possess the CSS\ structure.

First assume that such generators exist. It is not possible that
all generators consist of the same type of operators (all $X$ or
all $Z$) because single-qubit errors of the same type ($X$ or $Z$)
are then not correctable. Consider the possibility that there is
one generator of one type, say $X$, and four generators of the
other type, say $Z$. If the generator of type $X$ has an identity
acting on any qubit, say the first one, then the error $Z_{1}$
commutes with all generators. This error is not correctable unless
it belongs to the stabilizer. But if it belongs to the stabilizer,
the first qubit of the code must be fixed in the state
$|0\rangle$, which makes for a trivial code. The other possibility
is that the $X$-type generator has the form $g_{1}=XXXXXX$. But
then any combination of two $Z$-errors ($Z_{i}Z_{j}$) commutes
with it, and so they have to belong to the stabilizer. But there
are five independent such combinations of errors ($Z_{1}Z_{2}$,
$Z_{1}Z_{3}$, $Z_{1}Z_{4}$, $Z_{1}Z_{5}$, $Z_{1}Z_{6}$) and only
four generators of the $Z$ type. Therefore, it is impossible for
the code to have four generators of one type and one generator of
the other type.

The only possibility left is that there are two generators of one type, say
$X$, and three generators of the other type, say $Z$. The two $X$-type generators
should not both have identity acting on any given qubit because a $Z$ error on
that qubit commutes with all generators. Such an error cannot belong to the
stabilizer because it would again make for a trivial code. Specifically, we write the two $X$-type generators
($g_{1}$ and $g_{2}$) one above the other
\begin{equation}
\begin{matrix}
g_{1}\\
g_{2}
\end{matrix}
=
\begin{matrix}
- & - & - & - & - & -\\
- & - & - & - & - & -
\end{matrix},
\label{eq:matrix}
\end{equation}
where we leave the entries unspecified in the above equation,
but they are either $X$ or $I$. Both generators cannot have the column
\begin{equation}
\begin{matrix}
I\\
I
\end{matrix}
\nonumber
\end{equation}
in (\ref{eq:matrix}) because both generators
cannot have identities acting on the same qubit. Thus, only three different
columns can build up the generators in (\ref{eq:matrix}):
\begin{equation}
\begin{matrix}
I\\
X
\end{matrix}
\hspace{0.5cm},\hspace{0.5cm}
\begin{matrix}
X\\
I
\end{matrix}
\hspace{0.5cm},\hspace{0.5cm}
\begin{matrix}
X\\
X
\end{matrix}
\hspace{0.5cm}\nonumber
\end{equation}
We distinguish the following cases:
\begin{enumerate}
\item Each column appears twice.
\item One column appears three times, another column appears twice, and the
third column appears once.
\item One column appears three times and another column appears three times.
\item At least one column appears more than three times.
\end{enumerate}
If one and the same column appears on two different places, say qubit one and
qubit two as in the following example,
\begin{equation}
\begin{matrix}
g_{1}\\
g_{2}
\end{matrix}
=
\begin{matrix}
X & X & - & - & - & -\\
I & I & - & - & - & -
\end{matrix}
\hspace{0.5cm},
\end{equation}
then a pair of $Z$ errors on these qubits ($Z_{1}Z_{2}$) commutes with all
generators, and therefore belongs to the stabilizer.

In the first case considered above, there are three such pairs of errors,
which up to a relabeling of the qubits can be taken to be $Z_{1}Z_{2}$,
$Z_{3}Z_{4}$, $Z_{5}Z_{6}$. They can be used as stabilizer generators because
these operators are independent. But then the following pairs of single-qubit
$X$ errors commute with all generators: $X_{1}X_{2}$, $X_{3}X_{4}$,
$X_{5}X_{6}$. This possibility is ruled out because the latter cannot be part
of the stabilizer generators.

In the second case, up to a relabeling of the qubits, we have the following pairs of
$Z$ errors that commute with the stabilizer: $Z_{1}Z_{2}$, $Z_{1}Z_{3}$,
$Z_{2}Z_{3}$, $Z_{4}Z_{5}$. Only three of all four are independent, and they
can be taken to be stabilizer generators. But then all three generators of
$Z$-type have the identity acting on the sixth qubit, and therefore the error
$X_{6}$ is not correctable (and it cannot be a stabilizer generator because
it would make for a trivial code).

In the third case, the pairs $Z_{1}Z_{2}$, $Z_{1}Z_{3}$, $Z_{2}Z_{3}$, $Z_{4}Z_{5}$,
$Z_{4}Z_{6}$, $Z_{5}Z_{6}$ (up to a relabeling), four of which are
independent, commute with the stabilizer. But they cannot all belong to the
stabilizer because there are only three possible generators of the $Z$-type.

Finally, in the fourth case, we have three or more independent pairs of $Z$ errors
commuting with the stabilizer (for example $Z_{1}Z_{2}$, $Z_{1}Z_{3}$,
$Z_{1}Z_{4}$, which corresponds to the first four columns being identical). If
the independent pairs are more than three, then their number is more
than the possible number of generators. If they are exactly three, we can take
them as generators. But then $Z$-type generators act trivially upon two qubits,
and therefore $X$ errors on these qubits are not correctable. This last step
completes the proof.
\end{proof}

\section{Non-degenerate Six-Qubit CSS Entanglement-Assisted Quantum Code}

We detail the construction of a six-qubit
CSS entanglement-assisted quantum code in this section. We first
review the history of entanglement-assisted quantum coding and
discuss the operation of an entanglement-assisted code. We then
describe our construction. It turns out that the code we obtain is
equivalent to the Steane code \cite{Steane:96a} when including
Bob's qubit, and therefore is not a new code. It suggests,
however, a general rule for which we present a proof---every
$[[n,1,3]]$ code is equivalent to a $[[n-1,1,3;1]]$
entanglement-assisted code with any qubit serving as Bob's half of
the ebit. Even though our code is a trivial example of this rule,
it is instructive to present its derivation from the perspective
of the theory of entanglement-assisted codes.

\label{sec:bilal-mark}Bowen constructed an example of a quantum
error-correcting code that exploits shared entanglement between sender and
receiver \cite{PhysRevA.66.052313}. Brun, Devetak, and Hsieh later generalized
Bowen's example and developed the entanglement-assisted stabilizer formalism
\cite{arx2006brun,science2006brun}. This theory is an extension of the
standard stabilizer formalism and uses shared entanglement to formulate
stabilizer codes. Several references provide a review
\cite{arx2006brun,science2006brun,arxiv2007brun} and generalizations of the
theory to entanglement-assisted operator codes
\cite{isit2007brun,arxiv2007brun}, convolutional entanglement distillation
protocols \cite{arx2007wilde}, continuous-variable codes
\cite{arx2007wildeEA}, and entanglement-assisted quantum convolutional codes
\cite{arx2007wildeEAQCC, arx2008wildeUQCC}. Gilbert {\it et al.}~also generalized their \textquotedblleft
quantum computer condition\textquotedblright\ for fault tolerance to the
entanglement-assisted case \cite{arx2007gilbert}.  Entanglement-assisted
codes are a special case of ``correlation-assisted codes'', where Bob's qubit
is also allowed to be noisy. Such codes are in turn instances
of general linear quantum error-correcting codes \cite{SL:07}.

An entanglement-assisted quantum error-correcting code operates as
follows. A sender and receiver share $c$ ebits before
communication takes place. The sender possesses her half of the
$c$ ebits, $n-k-c$ ancilla qubits, and $k$ information qubits. She
performs an encoding unitary on her $n$\ qubits and sends them
over a noisy quantum communication channel. The receiver combines
his half of the $c$\ ebits with the $n$ encoded qubits and
performs measurements on all of the qubits to diagnose the errors
from the noisy channel. The generators corresponding to the
measurements on all of the qubits form a commuting set. The
generators thus form a valid stabilizer, they do not disturb the
encoded quantum information, and they learn only about the errors
from the noisy channel. The notation for such a code is
$[[n,k,d;c]]$, where $d$ is the distance of the code.

The typical assumption for an entanglement-assisted quantum code
is that noise does not affect Bob's half of the ebits because they
reside on the other side of a noisy quantum communication channel
between Alice and Bob. Our $[[6,1,3;1]]$ entanglement-assisted code
is globally equivalent to the $[[7,1,3]]$ Steane code
and thus corrects errors on Bob's side as well. From a
computational perspective, a code that corrects errors on all
qubits is more powerful than a code that does not. From the
perspective of the entanglement-assisted paradigm, however, this feature
is unnecessary and may result in decreased
error-correcting capabilities of the code with respect to errors
on Alice's side.

We construct our code using the parity check matrix of a classical
code. Consider the parity check matrix for the $\left[
7,4,3\right]$ Hamming
code:%
\begin{equation}
\left[
\begin{array}
[c]{ccccccc}%
1 & 0 & 0 & 1 & 0 & 1 & 1\\
0 & 1 & 0 & 1 & 1 & 0 & 1\\
0 & 0 & 1 & 0 & 1 & 1 & 1
\end{array}
\right]
\end{equation}
The Hamming code encodes four classical bits and corrects a single-bit error.
We remove one column of the above parity check matrix to form a new parity check matrix
$H$\ as follows:%
\begin{equation}
H=\left[
\begin{array}
[c]{cccccc}%
1 & 0 & 0 & 1 & 0 & 1\\
0 & 1 & 0 & 1 & 1 & 0\\
0 & 0 & 1 & 0 & 1 & 1
\end{array}
\right]  .
\end{equation}
The code corresponding to $H$\ encodes three bits and still corrects a
single-bit error. We begin constructing the stabilizer for an
entanglement-assisted quantum code by using the CSS construction
\cite{isit2007brun,arxiv2007brun}:%
\begin{equation}
\left[  \left.
\begin{array}
[c]{c}%
H\\
0
\end{array}
\right\vert
\begin{array}
[c]{c}%
0\\
H
\end{array}
\right]  .\label{eq:EA-six-qubit-binary}%
\end{equation}
The left side of the above matrix is the \textquotedblleft Z\textquotedblright%
\ side and the right side of the above matrix is the \textquotedblleft
X\textquotedblright\ side. The isomorphism between $n$-fold tensor products of
Pauli matrices and $n$-dimensional binary vectors gives a correspondence
between the matrix in (\ref{eq:EA-six-qubit-binary})\ and the set of
Pauli generators below \cite{thesis97gottesman,book2000mikeandike,arx2006brun}:%
\begin{equation}%
\begin{array}
[c]{cccccc}%
Z & I & I & Z & I & Z\\
I & Z & I & Z & Z & I\\
I & I & Z & I & Z & Z\\
X & I & I & X & I & X\\
I & X & I & X & X & I\\
I & I & X & I & X & X
\end{array}
\end{equation}
The above set of generators have good quantum error-correcting properties
because they correct an arbitrary single-qubit error. These properties follow
directly from the properties of the classical code. The problem with the above
generators is that they do not form a commuting set and thus do not correspond
to a valid quantum code. We use entanglement to resolve this problem by
employing the method outlined in
Ref.~\cite{arx2006brun,science2006brun,arxiv2007brun}.
\begin{table}[tbp] \centering
\begin{tabular}
[c]{ccc}%
\begin{tabular}
[c]{c|c|cccccc}
& Bob & \multicolumn{6}{|c}{Alice}\\\hline\hline
$g_{1}^{\prime}$ & $I$ & $I$ & $Z$ & $I$ & $I$ & $I$ & $I$\\
$g_{2}^{\prime}$ & $I$ & $I$ & $I$ & $Z$ & $I$ & $I$ & $I$\\
$g_{3}^{\prime}$ & $Z$ & $Z$ & $I$ & $I$ & $I$ & $I$ & $I$\\
$g_{4}^{\prime}$ & $I$ & $I$ & $I$ & $I$ & $Z$ & $I$ & $I$\\
$g_{5}^{\prime}$ & $I$ & $I$ & $I$ & $I$ & $I$ & $Z$ & $I$\\
$g_{6}^{\prime}$ & $X$ & $X$ & $I$ & $I$ & $I$ & $I$ & $I$\\\hline
$\overline{X}^{\prime}$ & $I$ & $I$ & $I$ & $I$ & $I$ & $I$ & $X$\\
$\overline{Z}^{\prime}$ & $I$ & $I$ & $I$ & $I$ & $I$ & $I$ & $Z$%
\\\hline\hline
\end{tabular}
& \quad\quad &
\begin{tabular}
[c]{c|c|cccccc}
& Bob & \multicolumn{6}{|c}{Alice}\\\hline\hline
$g_{1}$ & $I$ & $Z$ & $I$ & $Z$ & $Z$ & $Z$ & $I$\\
$g_{2}$ & $I$ & $Z$ & $Z$ & $I$ & $I$ & $Z$ & $Z$\\
$g_{3}$ & $Z$ & $Z$ & $I$ & $I$ & $Z$ & $I$ & $Z$\\
$g_{4}$ & $I$ & $X$ & $X$ & $I$ & $I$ & $X$ & $X$\\
$g_{5}$ & $I$ & $I$ & $X$ & $X$ & $X$ & $I$ & $X$\\
$g_{6}$ & $X$ & $X$ & $I$ & $I$ & $X$ & $I$ & $X$\\\hline
$\overline{X}$ & $I$ & $I$ & $I$ & $I$ & $X$ & $X$ & $X$\\
$\overline{Z}$ & $I$ & $I$ & $Z$ & $Z$ & $I$ & $Z$ & $I$\\\hline\hline
\end{tabular}
\\
(a) &  & (b)
\end{tabular}
\caption{(a) The generators and logical operators for the
unencoded state. Generators $g'_3$ and $g'_6$ indicate that Alice
and Bob share an ebit. Alice's half of the ebit is her first qubit
and Bob's qubit is the other half of the ebit. Generators $g'_1$,
$g'_2$, $g'_4$, and $g'_5$ indicate that Alice's second, third,
fourth, and fifth respective qubits are ancilla qubits in the
state $\left\vert 0 \right\rangle$. The unencoded logical
operators $\bar{X}'$ and $\bar{Z}'$ act on the sixth qubit and
indicate that the sixth qubit is the information qubit. (b) The
encoded generators and logical operators for the $[[6,1,3;1]]$
entanglement-assisted quantum error-correcting code.}\label{tbl:ea-613}%
\end{table}%

Three different but related methods determine the minimum number of ebits that
the entanglement-assisted quantum code requires:

\begin{enumerate}
\item Multiplication of the above generators with one another according to the
``symplectic
Gram-Schmidt orthogonalization algorithm''
forms a new set of generators \cite{arx2006brun,science2006brun}.
The error-correcting properties of the code are
invariant under these multiplications because the code is an additive code.
The resulting code has equivalent error-correcting properties and uses the
minimum number of ebits. We employ this technique in this work.

\item A slightly different algorithm in the appendix of Ref.~\cite{arx2007wilde}
determines the minimum number of ebits
required, the stabilizer measurements to perform, and the local encoding
unitary that Alice performs to rotate the unencoded state to the encoded state.
This algorithm is the most useful because it
\textquotedblleft kills three birds with one stone.\textquotedblright

\item The minimum number of ebits for a CSS\ entanglement-assisted code is equal to
the rank of $HH^{T}$ \cite{isit2007brun,arxiv2007brun,arx2008wildeOEA}. This simple formula is
useful if we are only concerned with computing the minimum number of ebits. It
does not determine the stabilizer generators or the encoding circuit.
Our code requires one ebit to form a valid
stabilizer code because the rank
of $HH^{T}$ for our code is equal to one.
\end{enumerate}

Table~\ref{tbl:ea-613}(b) gives the final form of the stabilizer
for our entanglement-assisted six-qubit code. We list both the
unencoded and the encoded generators for this code in
Table~\ref{tbl:ea-613}.

Our code uses one ebit shared between sender and receiver in the encoding
process. The sender performs a local encoding unitary that encodes one qubit
with the help of four ancilla qubits and one ebit.

The symplectic Gram-Schmidt algorithm yields a symplectic matrix that rotates
the unencoded symplectic vectors to the encoded symplectic vectors. The
symplectic matrix corresponds to an encoding unitary acting on the unencoded
quantum state \cite{arx2006brun,science2006brun}. This correspondence results
from the Stone-von Neumann Theorem and unifies the Schr\"{o}dinger and
Heisenberg pictures for quantum error correction \cite{eisert-2003-1}.

The symplectic Gram-Schmidt algorithm also determines the logical operators
for the code. Some of the vectors in the symplectic matrix that do not
correspond to a stabilizer generator are equivalent to the logical operators
for the code. It is straightforward to determine which symplectic vector
corresponds to which logical operator ($X$ or $Z$) by observing the action of
the symplectic matrix on vectors that correspond to the unencoded $X$ or $Z$
logical operators.

For our code, the symplectic matrix is as follows:%
\begin{equation}
\left[  \left.
\begin{array}
[c]{cccccc}%
1 & 0 & 0 & 1 & 0 & 1\\
1 & 0 & 1 & 1 & 1 & 0\\
1 & 1 & 0 & 0 & 1 & 1\\
0 & 0 & 0 & 0 & 0 & 0\\
0 & 0 & 0 & 0 & 0 & 0\\
0 & 1 & 1 & 0 & 1 & 0\\
0 & 0 & 0 & 0 & 0 & 0\\
0 & 0 & 0 & 0 & 0 & 0\\
0 & 0 & 0 & 0 & 0 & 0\\
0 & 0 & 0 & 1 & 0 & 1\\
0 & 1 & 0 & 1 & 0 & 1\\
0 & 0 & 0 & 0 & 0 & 0
\end{array}
\right\vert
\begin{array}
[c]{cccccc}%
0 & 0 & 0 & 0 & 0 & 0\\
0 & 0 & 0 & 0 & 0 & 0\\
0 & 0 & 0 & 0 & 0 & 0\\
1 & 1 & 0 & 0 & 1 & 1\\
0 & 1 & 1 & 1 & 0 & 1\\
0 & 0 & 0 & 0 & 0 & 0\\
1 & 0 & 0 & 1 & 0 & 1\\
0 & 0 & 1 & 1 & 1 & 1\\
0 & 0 & 1 & 0 & 1 & 0\\
0 & 0 & 0 & 0 & 0 & 0\\
0 & 0 & 0 & 0 & 0 & 0\\
0 & 0 & 0 & 1 & 1 & 1
\end{array}
\right]
\end{equation}
The index of the rows of the above matrix corresponds to the operators in the
unencoded stabilizer in Table~\ref{tbl:ea-613}(a). Therefore, the first five
rows correspond to the encoded $Z$ operators in the stabilizer and the sixth
row corresponds to the logical $\overline{Z}$ operator. As an example, we can
represent the unencoded logical $\overline{Z}$ operator in
Table~\ref{tbl:ea-613}(a) as the following binary vector:%
\begin{equation}
\left[  \left.
\begin{array}
[c]{cccccc}%
0 & 0 & 0 & 0 & 0 & 1
\end{array}
\right\vert
\begin{array}
[c]{cccccc}%
0 & 0 & 0 & 0 & 0 & 0
\end{array}
\right]  .
\end{equation}
Premultiplying the above matrix by the above row vector gives the binary form
for the encoded logical $\overline{Z}$ operator. We can then translate this
binary vector to a six-fold tensor product of Paulis equivalent to the
logical $\overline{Z}$ operator in Table~\ref{tbl:ea-613}(b). Using this same
idea, the first row of the above matrix corresponds to Alice's Paulis in
$g_{3}$, the second row to $g_{1}$, the third row to $g_{2}$, the fourth row
to $g_{4}$, the fifth row to $g_{5}$, and the seventh row to $g_{6}$. The last
six rows in the above matrix correspond to encoded $X$ operators and it is
only the last row that is interesting because it acts as a logical $X$ operator.

\begin{figure}
[ptb]
\begin{center}
\includegraphics{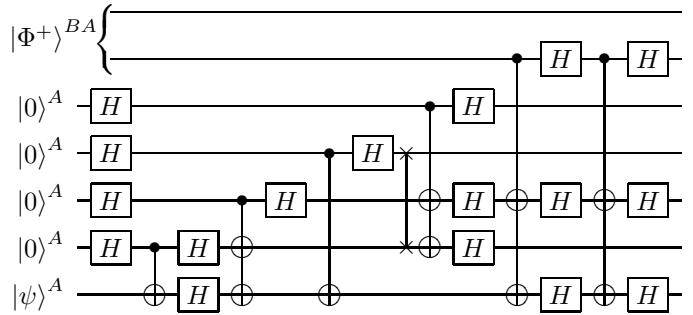}
\end{center}
\caption
{Encoding circuit for the [[6,1,3;1]] code.  The ``H'' gate is a Hadamard gate.}\label{fig-eacodecircuit}
\end{figure}%

Figure~\ref{fig-eacodecircuit} gives the encoding circuit for the
code.

We now detail the operations that give the equivalence of this
code to the seven-qubit Steane code. Consider the generators in
Table~\ref{tbl:ea-613}(b). Label the columns from left to right as
$1$, $2$, \ldots, $7$ where ``1'' corresponds to Bob's column.
Replace the generator $g_1$ by $g_1 g_2 g_3$, and the generator
$g_5$ by $g_5 g_6$. Switch the new generators $g_4$ and $g_5$.
Switch columns 2 and 3. Switch columns 1 and 5. Cyclically permute
the columns once so that 1 becomes 7, 2 becomes 1, 3 becomes 2,
..., 7 becomes 6. The resulting code is exactly the Steane code if
one reads it from right to left (i.e., up to the permutation $1
\leftrightarrow 7$, $2 \leftrightarrow 6$, $3 \leftrightarrow 5$).

Inspection of the encoded logical operators in
Table~\ref{tbl:ea-613}(b) reveals that Alice can perform logical
$\overline{X}$ and $\overline{Z}$ operations locally. Since the
CNOT\ has a transversal implementation for the Steane code, if
Alice and Bob possess two logical qubits each encoded with this
entanglement-assisted code, they can apply an encoded CNOT
transversally by the use of classical communication to coordinate
their actions. We point out, however, that the idea of computation
in the entanglement-assisted paradigm is not well motivated, since
if classical communication is allowed, Alice could send the
initial state to Bob and inform him of the operations that
need to be applied. An interesting open question is if there exist
codes that allow fault-tolerant computation on Alice's
side only.

From this example, we observe that some subset of the
entanglement-assisted codes correct errors on Bob's side.  This
phenomenon can be understood as an instance of the more general
correlation-assisted codes and linear quantum error-correction
theory detailed in Ref.~\cite{SL:07}. It may be useful from a
practical standpoint to determine which entanglement-assisted
codes satisfy this property. Here we provide an answer for the
case of single-error-correcting codes that use one bit of entanglement.

\begin{proposition}
Every $[[n,1,3]]$ code is equivalent
to a $[[n-1,1,3;1]]$ code with any qubit serving as Bob's half of
the ebit.
\end{proposition}

\begin{proof}
We prove this proposition by showing that any column in the table of
stabilizer generators for such a code can be reduced to the
standard form of Bob's column in an entanglement-assisted code
(containing exactly one $X$ and one $Z$ operator). Without loss of
generality, consider the column corresponding to the first qubit.
This column generally may contain $X$, $Y$, $Z$, or $I$ operators,
but if the code corrects any error on the first qubit, there must
be at least two different Pauli operators in this column. We can
reduce this column to the desired form as follows. Choose one of
the generators that contains $X$ on the first qubit, and replace
each of the other generators that contain an $X$ there by its
product with the chosen generator. Do the same for $Y$ and $Z$.
Thus we are left with at most one generator with $X$, one with $Y$
and one with $Z$. To eliminate $Y$, we replace it by its product
with the $X$ and $Z$ generators. If either $X$ or $Z$ is missing,
we replace the $Y$ generator with its product with the other
non-trivial generator.
\end{proof}

This result can be understood as a reflection of the fact that in
a code that corrects arbitrary single-qubit errors, every qubit is
maximally entangled with the rest and therefore can be thought of
as part of an ebit. The latter can also be seen to follow from the
property that every single-qubit error must send the code space to
an orthogonal subspace.

Note that for the case of $[[n,1,3;c]]$ codes with $c>1$, the
relation could be more complicated. If such a code corrects
an arbitrary single-qubit error, it is equivalent to an
$[[n+c,1,3]]$ code, but it is not obvious whether a $[[n+c,1,3]]$ code
can be interpreted as a $[[n,1,3;c]]$ code because the type of
entanglement that exists between $c$ qubits and the rest $n$
qubits may not be the same as that of $c$ e-bits.

\section{Non-existence of $[[n,1,3;1]]$ CSS codes for $n\leq 5$}
\label{sec:ogycss}

We now show that there does not exist a smaller entanglement-assisted
CSS code that uses only one ebit and corrects an arbitrary
single-qubit error on Alice's side. The proof is similar to that
for the non-existence of a $[[6,1,3]]$ CSS code.

\begin{proposition}
There does not exist an $[[n,1,3;1]]$ entanglement-assisted CSS
code for $n\leq 5$.
\end{proposition}

\begin{proof}
We being this proof by giving a dimensionality argument for the non-existence
of quantum codes (CSS or non-CSS) with $n < 4$.  This can be easily seen as follows.
Assume that the code is non-degenerate. There are $3n$ different single-qubit errors on
Alice's side, which means that there must exist $3n+1$ orthogonal
subspaces of dimension $2$ inside the entire $2^{n+1}$-dimensional
Hilbert space, i.e., $(3n+1)2\leq 2^{n+1}$. This is impossible for
$n<4$. Since for $n\leq3$ the number of generators is at most 3,
and two of the generators have to act non-trivially on Bob's side,
we can have degeneracy with respect to errors on Alice's side only
for $n=3$ with exactly one of the generators being equal to a pair
of errors on Alice's side. These two errors would be the only
indistinguishable single-qubit errors on Alice's side (no other
pair of errors on Alice's side can belong to the stabilizer),
which reduces the number of required orthogonal subspaces from
$3\times 3+1=10$ to 9. The required dimensions are $2\times 9=18$
and they cannot fit in the $2^4=16$-dimensional Hilbert space.

Suppose that there exists a $[[5,1,3;1]]$ CSS code. Its stabilizer
must have 5 generators ($S=\langle g_1,..., g_5\rangle$), each
consisting of only $X$ and $I$ operators or $Z$ and $I$ operators.
For an entanglement-assisted code, the generators must be of
the form
\begin{equation}
\begin{tabular}
[c]{cccccc|c}
$g_1=$ & $-$ & $-$ &$-$ & $-$ & $-$ & $X$\\
$g_2=$ & $-$ & $-$ &$-$ & $-$ & $-$ & $Z$\\
$g_3=$ & $-$ & $-$ &$-$ & $-$ & $-$ & $I$\\
$g_4=$ & $-$ & $-$ &$-$ & $-$ & $-$ & $I$ \\
$g_5=$ & $-$ & $-$ &$-$ & $-$ & $-$ & $I$
\end{tabular}
\end{equation}
where we have left the entries on Alice's side unspecified. The
set of correctable Pauli errors on Alice's side $\{E_j\in
\mathcal{P}_5\}$ (where $\mathcal{P}_5$ is the five-qubit Pauli
group) must satisfy $\{E_iE_j,S\}=0$ unless $E_i E_j \in S$, for
all $i,j=1,2,3,4,5$. All generators cannot be of the same type
($X$ or $Z$). The possibility that there is one generator of one
type, say $X$, and four generators of the other ($Z$) type, is
also ruled out because the $X$-type generator would have to be of
the form $g_1=XXXXX|X$ in order that every qubit is acted upon
non-trivially by at least one $X$ operator from the stabilizer.
This would mean, however, that any combination of two $Z$-errors
($Z_iZ_j$, $i,j=1,2,3,4,5$) would commute with the stabilizer, and
so it would have to belong to the stabilizer. There are four
independent such combinations of errors
($Z_1Z_2$,$Z_1Z_3$,$Z_1Z_4$,$Z_1Z_5$) which would have to be the
other four generators. But then there would be no possibility for
a $Z$ operator on Bob's side (as in $g_2$). Therefore, this is
impossible.

The only possibility is that there are 2 generators of one type,
say $X$, and 3 generators of the other type ($Z$). The two
$X$-type generators should not both have identity acting on any
given qubit on Alice's side because a $Z$ error on that qubit would
commute with all generators. Consider the following form for the two $X$-type generators:
\begin{equation}
\begin{tabular}
[c]{cccccc|c}
$g_1=$ & $-$ &$-$ & $-$ & $-$ & $-$ & $X$\\
$g_3=$ & $-$ &$-$ & $-$ & $-$ & $-$ & $I$
\end{tabular}
\end{equation}
There are three different columns that can fill the unspecified
entries in the above table:
\begin{equation}
\begin{matrix}
I\\
X
\end{matrix}\hspace{0.5cm}, \hspace{0.5cm}
\begin{matrix}
X\\
I
\end{matrix}\hspace{0.5cm}, \hspace{0.5cm}
\begin{matrix}
X\\
X
\end{matrix}\hspace{0.5cm}.
\notag
\end{equation}
We distinguish the following cases: two columns appear twice
and one column appears once, one column appears three times and
another column appears twice, one column appears three times and
each of the other columns appears once, at least one column
appears more than three times.

In the first case, up to relabeling of the qubits, we
distinguish the following possibilities:
\begin{equation}
\begin{tabular}
[c]{cccccc|c}
$g_1'=$ & $I$ & $I$ & $X$ & $X$ & $X$ & $X$\\
$g_3'=$ & $X$ & $X$ & $I$ & $I$ & $X$ & $I$
\end{tabular}\label{possibility1}
\end{equation}
\begin{equation}
\begin{tabular}
[c]{cccccc|c}
$g_1''=$ & $X$ &$X$ & $I$ & $I$ & $X$ & $X$\\
$g_3''=$ & $X$ &$X$ & $X$ & $X$ & $I$ & $I$
\end{tabular}\label{possibility2}
\end{equation}
\begin{equation}
\begin{tabular}
[c]{cccccc|c}
$g_1'''=$ & $X$ &$X$ & $X$ & $X$ & $I$ & $X$\\
$g_3'''=$ & $X$ &$X$ & $I$ & $I$ & $X$ & $I$
\end{tabular}\label{possibility3}
\end{equation}
For each possibility, the pairs of errors $Z_1Z_2$ and $Z_3Z_4$
commute with the stabilizer and therefore they would have to be
equal to the stabilizer generators $g_4$ and $g_5$. But the pairs
of errors $X_1X_2$ and $X_3X_4$ would commute with $g_1$, $g_3$,
$g_4$ and $g_5$. Since these errors do not belong to the
stabilizer, they would have to anti-commute with $g_3$. Therefore,
up to interchanging the first and second, or the third and fourth
qubits, the generator $g_2$ must have the form
\begin{equation}
\begin{tabular}
[c]{cccccc|c} $g_3=$ & $Z$ &$I$ & $Z$ & $I$ & $Z$ & $Z$
\end{tabular}.
\end{equation}
(Note that the fifth entry must be $Z$ because there must be at
least one generator that has a $Z$ acting on that qubit.) But it
can be verified that for each of the possibilities
\eqref{possibility1}, \eqref{possibility2} and
\eqref{possibility3}, $g_3$ anti-commutes with one of the $X$-type
generators. Therefore, the first case is impossible.

In the second case, one of the possible columns appears three
times and another column appears twice, e.g.,
\begin{equation}
\begin{tabular}
[c]{cccccc|c}
$g_1=$ & $X$ &$X$ & $X$ & $X$ & $X$ & $X$\\
$g_3=$ & $X$ &$X$ & $X$ & $I$ & $I$ & $I$
\end{tabular}
\end{equation}
In such a case we would have three independent pairs of $Z$ errors
($Z_1Z_2$, $Z_1Z_3$ and $Z_4Z_5$) which commute with the
stabilizer and therefore have to belong to it. But then there
would be no possibility for a $Z$ operator on Bob's side (the
generator $g_2$). Therefore this case is impossible.

In the third case, one column appears three times and each other
column appears once, as in

\begin{equation}
\begin{tabular}
[c]{cccccc|c}
$g_1=$ & $X$ &$X$ & $X$ & $X$ & $I$ & $X$\\
$g_3=$ & $X$ &$X$ & $X$ & $I$ & $X$ & $I$
\end{tabular}
\end{equation}
In this case, the pairs of errors $Z_1Z_2$ and $Z_1Z_3$ commute
with the stabilizer and must be equal to $g_4$ and $g_5$. But in
order for the fourth and fifth qubits to be each acted upon by at
least one $Z$ operator from the stabilizer, the generator $g_2$
would have to be of the form
\begin{equation}
\begin{tabular}
[c]{cccccc|c} $g_2=$ & $-$ &$-$ & $-$ & $Z$ & $Z$ & $Z$
\end{tabular}
\end{equation}
This means that the pair of errors $X_4X_5$ commutes with the
stabilizer, and since it is not part of the stabilizer, this case
is also impossible.

Finally, if one column appears more than three times, there would
be at least three independent pairs of $Z$ errors on Alice's side
which have to belong to the stabilizer. This leaves no possibility
for a $Z$ operator on Bob's side, i.e., this case is also ruled
out. Therefore, a $[[5,1,3;1]]$ CSS code does not exist.

In a similar way we can show that a $[[4,1,3;1]]$ CSS code does not
exist. Such a code would have 4 generators of the form
\begin{equation}
\begin{tabular}
[c]{ccccc|c}
$g_1=$ &  $-$ &$-$ & $-$ & $-$ & $X$\\
$g_2=$ &  $-$ &$-$ & $-$ & $-$ & $Z$\\
$g_3=$ &  $-$ &$-$ & $-$ & $-$ & $I$\\
$g_4=$ &  $-$ &$-$ & $-$ & $-$ & $I$
\end{tabular}
\end{equation}
The possibilities that all of the generators are of the same type,
or that one generator is of one type and the other three are of
the other type, are readily ruled out by arguments similar to
those for the $[[5,1,3;1]]$ code. The only possibility is two
$X$-type generators and two $Z$-type generators. The table of the
$X$-type generators
\begin{equation}
\begin{tabular}
[c]{ccccc|c}
$g_1=$  &$-$ & $-$ & $-$ & $-$ & $X$\\
$g_3=$  &$-$ & $-$ & $-$ & $-$ & $I$
\end{tabular}
\end{equation}
has to be filled by the same three columns we discussed before. As
we saw in our previous arguments, in the case when one column
appears three or more times there are at least two independent
pairs of errors on Alice's side which commute with the stabilizer.
These errors would have to belong to the stabilizer, but this
leaves no possibility for a $Z$ operator on Bob's side. In the
case when one column appears twice and another column appears
twice, the situation is analogous. The only other case is when one
column appears twice and each of the other two columns appears
once, as in
\begin{equation}
\begin{tabular}
[c]{ccccc|c}
$g_1=$  &$X$ & $X$ & $I$ & $X$ & $X$\\
$g_3=$  &$X$ & $X$ & $X$ & $I$ & $I$
\end{tabular}
\end{equation}
Since in this case the pair of errors $Z_1Z_2$ would commute with
the stabilizer, this pair would have to be equal to the generator
$g_4$. The third and fourth qubits each have to be acted upon by
at least one $Z$ operators from the stabilizer. Thus the generator
$g_2$ would have to have the form
\begin{equation}
\begin{tabular}
[c]{ccccc|c} $g_2=$  &$-$ & $-$ & $Z$ & $Z$ & $Z$
\end{tabular}.
\end{equation}
But then the pair $X_3X_4$ which does not belong to the stabilizer
would commute with all stabilizer generators. Therefore a
$[[4,1,3;1]]$ CSS code does not exist.\end{proof}

We point out that a $[[4,1,3;1]]$ non-CSS code was found in
Ref.~\cite{science2006brun}. This is the smallest
possible code that can encode one qubit with the use of only one
ebit, and at the same time correct an arbitrary single-qubit error
on Alice's side. Here we have identified an example of the smallest
possible CSS code with these characteristics.

\section{Summary and Conclusion}
We have discussed two different examples of a six-qubit code
and have included a subsystem construction for the degenerate six-qubit code.  Our proof
explains why a six-qubit CSS code does not exist and clarifies
earlier results in Ref.~\cite{ieee1998calderbank} based on a
search algorithm. An immediate corollary of our result is that the
seven-qubit Steane code is the smallest CSS code capable of
correcting an arbitrary single-qubit error. An interesting open
problem is to generalize this tight lower bound to the setting of
CSS codes with a higher distance. We expect that our proof
technique may be useful for this purpose.

Our first example is a degenerate six-qubit code that corrects an arbitrary 
single-qubit error. The presentation
of the encoding circuit and the operations required for a logical
$X$, $Z$, and CNOT\ should aid in the implementation and
operation of this code.  We have converted this code into
a subsystem code that is non-trivial and saturates the subsystem Singleton bound. Our six-qubit subsystem
code requires only four stabilizer measurements during the
recovery process.

Our second example is an entanglement-assisted $[[6,1,3;1]]$ CSS
code that is globally equivalent to the Steane seven-qubit code.
We have presented the construction of this code from a set of six
non-commuting generators on six qubits. We have further shown that
every $[[n,1,3]]$ code can be used as a $[[n-1,1,3;1]]$
entanglement-assisted code.

Based on the proof technique that we used for the earlier
six-qubit code, we have shown that the Steane code is an example
of the smallest entanglement-assisted code that possesses the
CSS structure and uses exactly one ebit.  Here too, an interesting
open problem is the generalization of this tight lower
bound to higher distance entanglement-assisted codes or to codes
that use more than one ebit.

\section{Acknowledgments}

B.A.S. acknowledges support from NSF grant No. CCF-0448658, M.M.W. from NSF grant No. CCF-0545845, and O.O. from NSF Grant No. CCF-0524822. D.A.L. was sponsored by NSF under grants CCF-0523675 and CCF-0726439, and by the United States Department of Defense. The views and conclusions contained in this document are those of the authors and should not be interpreted as representing the official policies, either expressly or implied, of the U.S. Government. The authors thank Todd Brun for useful discussions. 
 
\section{Appendix}

\subsection{Tables}
The tables in the appendix detail the error-correcting properties
of both of the $[[6,1,3]]$ codes. Each table lists all possible
pairs of single-qubit
errors and a corresponding generator of the code that anticommutes with the pair.%

\begin{table*}[tbp] \centering
\begin{tabular}
[c]{c|c||c|c||c|c||c|c||c|c||c|c||c|c||c|c||c|c||c|c||c|c}\hline\hline
Error & AG & Error & AG & Error & AG & Error & AG & Error & AG & Error & AG &
Error & AG & Error & AG & Error & AG & Error & AG & Error & AG\\\hline\hline
$X_1X_2$ & $h_1$ & $X_1X_3$ & $h_2$ & $X_1X_4$ & $h_1$ &
$X_1X_5$ & $h_1$ & $X_1X_6$ & $h_5$ & $X_1Y_2$ & $h_1$ &
$X_1Y_3$ & $h_2$ & $X_1Y_4$ & $h_2$ & $X_1Y_5$ & $h_3$ &
$X_1Y_6$ & $h_1$ & $X_1Z_2$ & $h_1$\\
$X_1Z_3$ & $h_1$ & $X_1Z_4$ & $h_2$ & $X_1Z_5$ & $h_3$ &
$X_1Z_6$ & $h_2$ & $X_2X_3$ & $h_1$ & $X_2X_4$ & $h_3$ &
$X_2X_5$ & $h_3$ & $X_2X_6$ & $h_1$ & $X_2Y_3$ & $h_1$ &
$X_2Y_4$ & $h_1$ & $X_2Y_5$ & $h_1$\\
$X_2Y_6$ & $h_2$ & $X_2Z_3$ & $h_5$ & $X_2Z_4$ & $h_1$ &
$X_2Z_5$ & $h_1$ & $X_2Z_6$ & $h_1$ & $X_3X_4$ & $h_1$ &
$X_3X_5$ & $h_1$ & $X_3X_6$ & $h_2$ & $X_3Y_4$ & $h_3$ &
$X_3Y_5$ & $h_2$ & $X_3Y_6$ & $h_1$\\
$X_3Z_4$ & $h_3$ & $X_3Z_5$ & $h_2$ & $X_3Z_6$ & $h_3$ &
$X_4X_5$ & $h_5$ & $X_4X_6$ & $h_1$ & $X_4Y_5$ & $h_1$ &
$X_4Y_6$ & $h_2$ & $X_4Z_5$ & $h_1$ & $X_4Z_6$ & $h_1$ &
$X_5X_6$ & $h_1$ & $X_5Y_6$ & $h_2$\\
$X_5Z_6$ & $h_1$ & $Y_1X_2$ & $h_2$ & $Y_1X_3$ & $h_1$ &
$Y_1X_4$ & $h_2$ & $Y_1X_5$ & $h_2$ & $Y_1X_6$ & $h_1$ &
$Y_1Y_2$ & $h_3$ & $Y_1Y_3$ & $h_1$ & $Y_1Y_4$ & $h_1$ &
$Y_1Y_5$ & $h_1$ & $Y_1Y_6$ & $h_3$\\
$Y_1Z_2$ & $h_5$ & $Y_1Z_3$ & $h_2$ & $Y_1Z_4$ & $h_1$ &
$Y_1Z_5$ & $h_1$ & $Y_1Z_6$ & $h_1$ & $Y_2X_3$ & $h_1$ &
$y_2X_4$ & $h_2$ & $Y_2X_5$ & $h_2$ & $Y_2X_6$ & $h_1$ &
$Y_2Y_3$ & $h_1$ & $Y_2Y_4$ & $h_1$\\
$Y_2Y_5$ & $h_1$ & $Y_2Y_6$ & $h_5$ & $Y_2Z_3$ & $h_5$ &
$Y_2Z_4$ & $h_1$ & $Y_2Z_5$ & $h_1$ & $Y_2Z_6$ & $h_1$ &
$Y_3X_4$ & $h_1$ & $Y_3X_5$ & $h_1$ & $Y_3X_6$ & $h_2$ &
$Y_3Y_4$ & $h_5$ & $Y_3Y_5$ & $h_2$\\
$Y_3Y_6$ & $h_1$ & $Y_3Z_4$ & $h_5$ & $Y_3Z_5$ & $h_2$ &
$Y_3Z_6$ & $h_5$ & $Y_4X_5$ & $h_1$ & $Y_4X_6$ & $h_2$ &
$Y_4Y_5$ & $h_2$ & $Y_4Y_6$ & $h_1$ & $Y_4Z_5$ & $h_2$ &
$Y_4Z_6$ & $h_4$ & $Y_5X_6$ & $h_3$\\
$Y_5Y_6$ & $h_1$ & $Y_5Z_6$ & $h_2$ & $Z_1X_2$ & $h_1$ &
$Z_1X_3$ & $h_5$ & $Z_1X_4$ & $h_1$ & $Z_1X_5$ & $h_1$ &
$Z_1X_6$ & $h_2$ & $Z_1Y_2$ & $h_1$ & $Z_1Y_3$ & $h_3$ &
$Z_1Y_4$ & $h_3$ & $Z_1Y_5$ & $h_2$\\
$Z_1Y_6$ & $h_1$ & $Z_1Z_2$ & $h_1$ & $Z_1Z_3$ & $h_1$ &
$Z_1Z_4$ & $h_3$ & $Z_1Z_5$ & $h_2$ & $Z_1Z_6$ & $h_3$ &
$Z_2X_3$ & $h_1$ & $Z_2X_4$ & $h_2$ & $Z_2X_5$ & $h_2$ &
$Z_2X_6$ & $h_1$ & $Z_2Y_3$ & $h_1$\\
$Z_2Y_4$ & $h_1$ & $Z_2Y_5$ & $h_1$ & $Z_2Y_6$ & $h_3$ &
$Z_2Z_3$ & $h_2$ & $Z_2Z_4$ & $h_1$ & $Z_2Z_5$ & $h_1$ &
$Z_2Z_6$ & $h_1$ & $Z_3X_4$ & $h_3$ & $Z_3X_5$ & $h_3$ &
$Z_3X_6$ & $h_1$ & $Z_3Y_4$ & $h_1$\\
$Z_3Y_5$ & $h_1$ & $Z_3Y_6$ & $h_2$ & $Z_3Z_4$ & $h_1$ &
$Z_3Z_5$ & $h_1$ & $Z_3Z_6$ & $h_1$ & $Z_4X_5$ & $h_1$ &
$Z_4X_6$ & $h_2$ & $Z_4Y_5$ & $h_2$ & $Z_4Y_6$ & $h_1$ &
$Z_4Z_5$ & $h_2$ & $Z_4Z_6$ & $h_4$\\
$Z_5X_6$ & $h_3$ & $Z_5Y_6$ & $h_1$ & $Z_5Z_6$ & $h_2$ &
$X_1$ & $h_1$ & $X_2$ & $h_3$ & $X_3$ & $h_1$ &
$X_4$ & $h_4$ & $X_5$ & $h_5$ & $X_6$ & $h_1$ &
$Y_1$ & $h_2$ & $Y_2$ & $h_2$  \\
$Y_3$ & $h_3$ & $Y_4$ & $h_3$ & $Y_5$ & $h_3$ &
$Y_6$ & $h_3$ & $Z_1$ & $h_1$ & $Z_2$ & $h_2$ &
$Z_3$ & $h_3$ & $Z_4$ & $h_3$ & $Z_5$ & $h_3$ &
$Z_6$ & $h_1$ \\ \hline\hline
\end{tabular}
\caption{Distinct pairs of single-qubit Pauli errors for the $[[6,1,3]]$
quantum code. Each double-lined column lists a pair of single-qubit
errors and a corresponding anticommuting generator (AG) for the code.  $X_4$ and $Z_4Z_6$ lie in the gauge subgroup $H$.}\label{table:sixonethree1}%
\end{table*}%

\subsection{Entanglement-Assisted Encoding Circuit}
Here we detail an algorithm that generates the encoding circuit for the $[[6,1,3;1]]$ code.
We follow the recipe outlined in the appendix of Ref.~\cite{arx2007wilde}.
We begin by first converting the stabilizer generators in Table~\ref{tbl:ea-613}(b) into a
binary form which we refer to as a $Z|X$ matrix.  We obtain the the left $Z$ submatrix
by inserting a ``1'' wherever we see a $Z$ in the stabilizer generators.  We obtain the
$X$ submatrix by inserting a ``1'' wherever we see a corresponding $X$
in the stabilizer generator.  If there is a $Y$ in the generator, we insert a ``1''
in the corresponding row and column of both the $Z$ and $X$ submatrices.

The idea is to convert (\ref{appendixeqn1}) to (\ref{appendixeqn17}) through a series of row and column operations.
The binary form of the matrix in (\ref{appendixeqn1}) corresponds to the stabilizer generators in Table~\ref{tbl:ea-613}(b)
by employing the Pauli-to-binary isomorphism outlined in Ref.~\cite{book2000mikeandike}.
We can use CNOT, Hadamard, Phase, and SWAP gates.
\begin{enumerate}
\item When we apply a CNOT gate from qubit $i$ to qubit $j$, it adds column $i$ to
column $j$ in the $X$ submatrix, and in the $Z$ submatrix it adds column $j$ to column $i$.
\item A Hadamard on qubit $i$ swaps
column $i$ in the $Z$ submatrix with column $i$ in the $X$ submatrix.
\item A Phase gate on qubit
$i$ adds column $i$ in the $X$ submatrix to column $i$ in the $Z$ submatrix.
\item When we apply a
SWAP gate from qubit $i$ to qubit $j$, we exchange column $i$ with column $j$ in $Z$ submatrix
and column $i$ and column $j$ in the $X$ submatrix.
\end{enumerate}
Row operations do not change the error-correcting properties of the code.  They do not
cost us in terms of gates.  They are also crucial in determining the minimum number of ebits for the code.

\begin{equation}
\left[  \left.
\begin{array}
[c]{cccccc}
1 & 0 & 0 & 1 & 0 & 1 \\
1 & 0 & 1 & 1 & 1 & 0 \\
1 & 1 & 0 & 0 & 1 & 1 \\
0 & 0 & 0 & 0 & 0 & 0 \\
0 & 0 & 0 & 0 & 0 & 0 \\
0 & 0 & 0 & 0 & 0 & 0
\end{array}
\right\vert
\begin{array}
[c]{cccccc}
0 & 0 & 0 & 0 & 0 & 0 \\
0 & 0 & 0 & 0 & 0 & 0 \\
0 & 0 & 0 & 0 & 0 & 0 \\
1 & 1 & 0 & 0 & 1 & 1 \\
1 & 0 & 1 & 1 & 1 & 0 \\
1 & 0 & 0 & 1 & 0 & 1
\end{array}
\right]
\label{appendixeqn1}
\end{equation}
We begin the algorithm by computing the symplectic product \cite{arx2006brun} between the various rows of the matrix.
The first row is symplectically orthogonal to the second row.  Moreover, it is symplectically orthogonal
to all the rows except row six.  So we swap the second row with the sixth row.
\begin{equation}
\left[  \left.
\begin{array}
[c]{cccccc}
1 & 0 & 0 & 1 & 0 & 1 \\
0 & 0 & 0 & 0 & 0 & 0 \\
1 & 1 & 0 & 0 & 1 & 1 \\
0 & 0 & 0 & 0 & 0 & 0 \\
0 & 0 & 0 & 0 & 0 & 0 \\
1 & 0 & 1 & 1 & 1 & 0
\end{array}
\right\vert
\begin{array}
[c]{cccccc}
0 & 0 & 0 & 0 & 0 & 0 \\
1 & 0 & 0 & 1 & 0 & 1 \\
0 & 0 & 0 & 0 & 0 & 0 \\
1 & 1 & 0 & 0 & 1 & 1 \\
1 & 0 & 1 & 1 & 1 & 0 \\
0 & 0 & 0 & 0 & 0 & 0
\end{array}
\right]
\label{appendixeqn2}
\end{equation}
Now apply Hadamard gates to qubits, one, four and six.
This operation swaps the columns one, four and six on the $Z$ side with columns
one, four and six on the $X$ side.
\begin{equation}
\left[  \left.
\begin{array}
[c]{cccccc}
0 & 0 & 0 & 0 & 0 & 0 \\
1 & 0 & 0 & 1 & 0 & 1 \\
0 & 1 & 0 & 0 & 1 & 0 \\
1 & 0 & 0 & 0 & 0 & 1 \\
1 & 0 & 0 & 1 & 0 & 0 \\
0 & 0 & 1 & 0 & 1 & 0
\end{array}
\right\vert
\begin{array}
[c]{cccccc}
1 & 0 & 0 & 1 & 0 & 1 \\
0 & 0 & 0 & 0 & 0 & 0 \\
1 & 0 & 0 & 0 & 0 & 1 \\
0 & 1 & 0 & 0 & 1 & 0 \\
0 & 0 & 1 & 0 & 1 & 0 \\
1 & 0 & 0 & 1 & 0 & 0
\end{array}
\right]
\label{appendixeqn3}
\end{equation}
Apply a CNOT from qubit one to qubit four and a CNOT from qubit one to qubit six.
This operation adds column one to four and column one to column six on the $X$ side.  On the $Z$
side of the matrix, the CNOT operation adds column four to column one and column six to column one.
\begin{equation}
\left[  \left.
\begin{array}
[c]{cccccc}
0 & 0 & 0 & 0 & 0 & 0 \\
1 & 0 & 0 & 1 & 0 & 1 \\
0 & 1 & 0 & 0 & 1 & 0 \\
0 & 0 & 0 & 0 & 0 & 1 \\
0 & 0 & 0 & 1 & 0 & 0 \\
0 & 0 & 1 & 0 & 1 & 0
\end{array}
\right\vert
\begin{array}
[c]{cccccc}
1 & 0 & 0 & 0 & 0 & 0 \\
0 & 0 & 0 & 0 & 0 & 0 \\
1 & 0 & 0 & 1 & 0 & 0 \\
0 & 1 & 0 & 0 & 1 & 0 \\
0 & 0 & 1 & 0 & 1 & 0 \\
1 & 0 & 0 & 0 & 0 & 1
\end{array}
\right]
\label{appendixeqn4}
\end{equation}
Now apply a Hadamard gate on qubit one.
\begin{equation}
\left[  \left.
\begin{array}
[c]{cccccc}
1 & 0 & 0 & 0 & 0 & 0 \\
0 & 0 & 0 & 1 & 0 & 1 \\
0 & 1 & 0 & 0 & 1 & 0 \\
0 & 0 & 0 & 0 & 0 & 1 \\
0 & 0 & 0 & 1 & 0 & 0 \\
0 & 0 & 1 & 0 & 1 & 0
\end{array}
\right\vert
\begin{array}
[c]{cccccc}
0 & 0 & 0 & 0 & 0 & 0 \\
1 & 0 & 0 & 0 & 0 & 0 \\
0 & 0 & 0 & 1 & 0 & 0 \\
0 & 1 & 0 & 0 & 1 & 0 \\
0 & 0 & 1 & 0 & 1 & 0 \\
0 & 0 & 0 & 0 & 0 & 1
\end{array}
\right]
\label{appendixeqn5}
\end{equation}
Apply a Hadamard gate on qubit four and qubit six.  This operation swaps columns four and six on
$Z$ side with respective columns on the $X$ side.
\begin{equation}
\left[  \left.
\begin{array}
[c]{cccccc}
1 & 0 & 0 & 0 & 0 & 0 \\
0 & 0 & 0 & 0 & 0 & 0 \\
0 & 1 & 0 & 1 & 1 & 0 \\
0 & 0 & 0 & 0 & 0 & 0 \\
0 & 0 & 0 & 0 & 0 & 0 \\
0 & 0 & 1 & 0 & 1 & 1
\end{array}
\right\vert
\begin{array}
[c]{cccccc}
0 & 0 & 0 & 0 & 0 & 0 \\
1 & 0 & 0 & 1 & 0 & 1 \\
0 & 0 & 0 & 0 & 0 & 0 \\
0 & 1 & 0 & 0 & 1 & 1 \\
0 & 0 & 1 & 1 & 1 & 0 \\
0 & 0 & 0 & 0 & 0 & 0
\end{array}
\right]
\label{appendixeqn6}
\end{equation}
Finally, we apply a CNOT gate from qubit one to qubit four and another CNOT gate
from qubit one to qubit six.
\begin{equation}
\left[  \left.
\begin{array}
[c]{cccccc}
1 & 0 & 0 & 0 & 0 & 0 \\
0 & 0 & 0 & 0 & 0 & 0 \\
0 & 1 & 0 & 1 & 1 & 0 \\
0 & 0 & 0 & 0 & 0 & 0 \\
0 & 0 & 0 & 0 & 0 & 0 \\
0 & 0 & 1 & 0 & 1 & 1
\end{array}
\right\vert
\begin{array}
[c]{cccccc}
0 & 0 & 0 & 0 & 0 & 0 \\
1 & 0 & 0 & 0 & 0 & 0 \\
0 & 0 & 0 & 0 & 0 & 0 \\
0 & 1 & 0 & 0 & 1 & 1 \\
0 & 0 & 1 & 1 & 1 & 0 \\
0 & 0 & 0 & 0 & 0 & 0
\end{array}
\right]
\label{appendixeqn7}
\end{equation}
At this point we are done processing qubit one and qubit two.  We now proceed to manipulate columns
two through six on the $Z$ and $X$ side.  We apply a Hadamard gate on qubit two, four and five.
\begin{equation}
\left[  \left.
\begin{array}
[c]{cccccc}
1 & 0 & 0 & 0 & 0 & 0 \\
0 & 0 & 0 & 0 & 0 & 0 \\
0 & 0 & 0 & 0 & 0 & 0 \\
0 & 1 & 0 & 0 & 1 & 0 \\
0 & 0 & 0 & 1 & 1 & 0 \\
0 & 0 & 1 & 0 & 0 & 1
\end{array}
\right\vert
\begin{array}
[c]{cccccc}
0 & 0 & 0 & 0 & 0 & 0 \\
1 & 0 & 0 & 0 & 0 & 0 \\
0 & 1 & 0 & 1 & 1 & 0 \\
0 & 0 & 0 & 0 & 0 & 1 \\
0 & 0 & 1 & 0 & 0 & 0 \\
0 & 0 & 0 & 0 & 1 & 0
\end{array}
\right]
\label{appendixeqn8}
\end{equation}
Perform a CNOT gate from qubit two to qubit four and from qubit two to qubit five.
\begin{equation}
\left[  \left.
\begin{array}
[c]{cccccc}
1 & 0 & 0 & 0 & 0 & 0 \\
0 & 0 & 0 & 0 & 0 & 0 \\
0 & 0 & 0 & 0 & 0 & 0 \\
0 & 0 & 0 & 0 & 1 & 0 \\
0 & 0 & 0 & 1 & 1 & 0 \\
0 & 0 & 1 & 0 & 0 & 1
\end{array}
\right\vert
\begin{array}
[c]{cccccc}
0 & 0 & 0 & 0 & 0 & 0 \\
1 & 0 & 0 & 0 & 0 & 0 \\
0 & 1 & 0 & 0 & 0 & 0 \\
0 & 0 & 0 & 0 & 0 & 1 \\
0 & 0 & 1 & 0 & 0 & 0 \\
0 & 0 & 0 & 0 & 1 & 0
\end{array}
\right]
\label{appendixeqn9}
\end{equation}
Perform a Hadamard on qubit two.
\begin{equation}
\left[  \left.
\begin{array}
[c]{cccccc}
1 & 0 & 0 & 0 & 0 & 0 \\
0 & 0 & 0 & 0 & 0 & 0 \\
0 & 1 & 0 & 0 & 0 & 0 \\
0 & 0 & 0 & 0 & 1 & 0 \\
0 & 0 & 0 & 1 & 1 & 0 \\
0 & 0 & 1 & 0 & 0 & 1
\end{array}
\right\vert
\begin{array}
[c]{cccccc}
0 & 0 & 0 & 0 & 0 & 0 \\
1 & 0 & 0 & 0 & 0 & 0 \\
0 & 0 & 0 & 0 & 0 & 0 \\
0 & 0 & 0 & 0 & 0 & 1 \\
0 & 0 & 1 & 0 & 0 & 0 \\
0 & 0 & 0 & 0 & 1 & 0
\end{array}
\right]
\label{appendixeqn10}
\end{equation}
We have processed qubit three.  Now look at the submatrix from columns three to
six on the $Z$ and $X$ side.  Perform a SWAP gate between qubit three and qubit five.
This operation swaps column three with five in the $Z$ submatrix and column three and
five in the $X$ submatrix.
\begin{equation}
\left[  \left.
\begin{array}
[c]{cccccc}
1 & 0 & 0 & 0 & 0 & 0 \\
0 & 0 & 0 & 0 & 0 & 0 \\
0 & 1 & 0 & 0 & 0 & 0 \\
0 & 0 & 1 & 0 & 0 & 0 \\
0 & 0 & 0 & 1 & 1 & 0 \\
0 & 0 & 1 & 0 & 0 & 1
\end{array}
\right\vert
\begin{array}
[c]{cccccc}
0 & 0 & 0 & 0 & 0 & 0 \\
1 & 0 & 0 & 0 & 0 & 0 \\
0 & 0 & 0 & 0 & 0 & 0 \\
0 & 0 & 0 & 0 & 0 & 1 \\
0 & 0 & 0 & 0 & 1 & 0 \\
0 & 0 & 1 & 0 & 0 & 0
\end{array}
\right]
\label{appendixeqn11}
\end{equation}
Perform a Hadamard gate on qubit three, followed by a CNOT gate from qubit three to qubit six,
and another Hadamard on qubit three.
\begin{equation}
\left[  \left.
\begin{array}
[c]{cccccc}
1 & 0 & 0 & 0 & 0 & 0 \\
0 & 0 & 0 & 0 & 0 & 0 \\
0 & 1 & 0 & 0 & 0 & 0 \\
0 & 0 & 1 & 0 & 0 & 0 \\
0 & 0 & 1 & 1 & 0 & 0 \\
0 & 0 & 0 & 0 & 1 & 1
\end{array}
\right\vert
\begin{array}
[c]{cccccc}
0 & 0 & 0 & 0 & 0 & 0 \\
1 & 0 & 0 & 0 & 0 & 0 \\
0 & 0 & 0 & 0 & 0 & 0 \\
0 & 0 & 0 & 0 & 0 & 0 \\
0 & 0 & 0 & 0 & 1 & 1 \\
0 & 0 & 0 & 0 & 0 & 0
\end{array}
\right]
\label{appendixeqn12}
\end{equation}
Add row four to five.
\begin{equation}
\left[  \left.
\begin{array}
[c]{cccccc}
1 & 0 & 0 & 0 & 0 & 0 \\
0 & 0 & 0 & 0 & 0 & 0 \\
0 & 1 & 0 & 0 & 0 & 0 \\
0 & 0 & 1 & 0 & 0 & 0 \\
0 & 0 & 0 & 1 & 0 & 0 \\
0 & 0 & 0 & 0 & 1 & 1
\end{array}
\right\vert
\begin{array}
[c]{cccccc}
0 & 0 & 0 & 0 & 0 & 0 \\
1 & 0 & 0 & 0 & 0 & 0 \\
0 & 0 & 0 & 0 & 0 & 0 \\
0 & 0 & 0 & 0 & 0 & 0 \\
0 & 0 & 0 & 0 & 1 & 1 \\
0 & 0 & 0 & 0 & 0 & 0
\end{array}
\right]
\label{appendixeqn13}
\end{equation}
We have completed processing qubit four. Now focus on columns four to six. Apply a Hadamard gate on qubit four,
followed by CNOT gate from qubit four to qubit five, and again from qubit four to qubit six.
\begin{equation}
\left[  \left.
\begin{array}
[c]{cccccc}
1 & 0 & 0 & 0 & 0 & 0 \\
0 & 0 & 0 & 0 & 0 & 0 \\
0 & 1 & 0 & 0 & 0 & 0 \\
0 & 0 & 1 & 0 & 0 & 0 \\
0 & 0 & 0 & 0 & 0 & 0 \\
0 & 0 & 0 & 0 & 1 & 1
\end{array}
\right\vert
\begin{array}
[c]{cccccc}
0 & 0 & 0 & 0 & 0 & 0 \\
1 & 0 & 0 & 0 & 0 & 0 \\
0 & 0 & 0 & 0 & 0 & 0 \\
0 & 0 & 0 & 0 & 0 & 0 \\
0 & 0 & 0 & 1 & 0 & 0 \\
0 & 0 & 0 & 0 & 0 & 0
\end{array}
\right]
\label{appendixeqn14}
\end{equation}
Perform a Hadamard gate on qubit four.
\begin{equation}
\left[  \left.
\begin{array}
[c]{cccccc}
1 & 0 & 0 & 0 & 0 & 0 \\
0 & 0 & 0 & 0 & 0 & 0 \\
0 & 1 & 0 & 0 & 0 & 0 \\
0 & 0 & 1 & 0 & 0 & 0 \\
0 & 0 & 0 & 1 & 0 & 0 \\
0 & 0 & 0 & 0 & 1 & 1
\end{array}
\right\vert
\begin{array}
[c]{cccccc}
0 & 0 & 0 & 0 & 0 & 0 \\
1 & 0 & 0 & 0 & 0 & 0 \\
0 & 0 & 0 & 0 & 0 & 0 \\
0 & 0 & 0 & 0 & 0 & 0 \\
0 & 0 & 0 & 0 & 0 & 0 \\
0 & 0 & 0 & 0 & 0 & 0
\end{array}
\right]
\label{appendixeqn15}
\end{equation}
Now look at columns five and six.  Apply a Hadamard gate on qubit five and qubit six, followed by a
CNOT gate from qubit five to qubit six.
\begin{equation}
\left[  \left.
\begin{array}
[c]{cccccc}
1 & 0 & 0 & 0 & 0 & 0 \\
0 & 0 & 0 & 0 & 0 & 0 \\
0 & 1 & 0 & 0 & 0 & 0 \\
0 & 0 & 1 & 0 & 0 & 0 \\
0 & 0 & 0 & 1 & 0 & 0 \\
0 & 0 & 0 & 0 & 0 & 0
\end{array}
\right\vert
\begin{array}
[c]{cccccc}
0 & 0 & 0 & 0 & 0 & 0 \\
1 & 0 & 0 & 0 & 0 & 0 \\
0 & 0 & 0 & 0 & 0 & 0 \\
0 & 0 & 0 & 0 & 0 & 0 \\
0 & 0 & 0 & 0 & 0 & 0 \\
0 & 0 & 0 & 0 & 1 & 0
\end{array}
\right]
\label{appendixeqn16}
\end{equation}
Perform a Hadamard on qubit five.
\begin{equation}
\left[  \left.
\begin{array}
[c]{cccccc}
1 & 0 & 0 & 0 & 0 & 0 \\
0 & 0 & 0 & 0 & 0 & 0 \\
0 & 1 & 0 & 0 & 0 & 0 \\
0 & 0 & 1 & 0 & 0 & 0 \\
0 & 0 & 0 & 1 & 0 & 0 \\
0 & 0 & 0 & 0 & 1 & 0
\end{array}
\right\vert
\begin{array}
[c]{cccccc}
0 & 0 & 0 & 0 & 0 & 0 \\
1 & 0 & 0 & 0 & 0 & 0 \\
0 & 0 & 0 & 0 & 0 & 0 \\
0 & 0 & 0 & 0 & 0 & 0 \\
0 & 0 & 0 & 0 & 0 & 0 \\
0 & 0 & 0 & 0 & 0 & 0
\end{array}
\right]
\label{appendixeqn17}
\end{equation}
We have finally obtained a binary matrix that corresponds to the canonical stabilizer
generators in Table~\ref{tbl:ea-613}(a).  Figure~\ref{fig-eacodecircuit} gives the encoding
circuit for the all the quantum operations that we performed above.  Multiplying the above
operations in reverse takes us from the unencoded canonical stabilizers to the encoded ones.

\bibliographystyle{apsrev}
\bibliography{sixonethree}

\end{document}